\documentclass[nature,aps,amsfonts,floatfix,superscriptaddress,longbibliography,nofootinbib,twocolumn]{revtex4-2}
\usepackage[utf8]{inputenc}
\usepackage{amsmath}    
\usepackage{mathbbol}
\usepackage{graphicx}   
\usepackage{verbatim}   
\usepackage{color}      
\usepackage{subfigure}  
\usepackage{hyperref} 

\usepackage{siunitx}
\usepackage{physics}
\usepackage{soul}
\usepackage{enumerate}
\usepackage{empheq}
\usepackage{multirow}

\begin{document}

\title{Driving superconducting qubits into chaos}

\author{Jorge Ch\'avez-Carlos}
\affiliation{Department of Physics, University of Connecticut, Storrs, Connecticut 06269, USA}

\author{Miguel A. Prado Reynoso} 
\affiliation{Department of Physics, University of Connecticut, Storrs, Connecticut 06269, USA}

\author{Ignacio Garc\'ia-Mata}
\affiliation{Also at Instituto de Investigaciones Físicas de Mar del Plata (IFIMAR), Facultad de Ciencias Exactas y Naturales,
Universidad Nacional de Mar del Plata, CONICET, 7600 Mar del Plata, Argentina}

\author{Victor S. Batista}
\affiliation{Department of Chemistry, Yale University, 
P.O. Box 208107, New Haven, Connecticut 06520-8107, USA}

\author{Francisco P\'erez-Bernal}
\affiliation{Departamento de Ciencias Integradas y Centro de Estudios Avanzados en F\'isica, Matem\'aticas y Computaci\'on, Universidad de Huelva, Huelva 21071, Spain}
\altaffiliation{Instituto Carlos I de Física Teórica y Computacional,
  Universidad de Granada, Granada 18071, SPAIN}

\author{Diego A. Wisniacki}
\affiliation{Departamento de F\'isica ``J. J. Giambiagi'' and IFIBA, FCEyN,
Universidad de Buenos Aires, 1428 Buenos Aires, Argentina}  

\author{Lea F. Santos}
\affiliation{Department of Physics, University of Connecticut, Storrs, Connecticut 06269, USA.}

\begin{abstract}
Kerr parametric oscillators are potential building blocks for fault-tolerant quantum computers. They can stabilize Kerr-cat qubits, which offer advantages toward the encoding and manipulation of error-protected quantum information. The recent realization of Kerr-cat qubits made use of the nonlinearity of the SNAIL transmon superconducting circuit and a squeezing drive. Increasing nonlinearities can enable faster gate times, but, as shown here, can also induce chaos and melt the qubit away.  We determine the region of validity of the Kerr-cat qubit and discuss how its disintegration could be experimentally detected. The danger zone for parametric quantum computation is also a potential playground for investigating quantum chaos with driven superconducting circuits.  
\end{abstract}

\maketitle


\noindent 
Decoherence is a familiar threat to quantum technologies. A resourceful way to protect quantum information against decoherence processes that act locally is to encode it nonlocally in the form of superpositions of coherent states~\cite{Mirrahimi2014}. These Schr\"odinger cat states~\cite{Haroche2013,Wineland2013,Leghtas2015} are the logical states of the so-called Kerr-cat qubit, which can be generated with driven Kerr parametric oscillators~\cite{Puri2017,Goto2018,Goto2019,Kwon2022,He2023}, as those experimentally realized in superconducting  circuits~\cite{Grimm2020}. However, the present work warns against the potential development of chaos if the parameters of the oscillators are pushed beyond a threshold. As explained here, the onset of local chaos can disintegrate the Kerr-cat qubit.

To stabilize Schr\"odinger cat states, the experiments combine Kerr nonlinearity and a squeezing (two-photon) drive. The nonlinear oscillator is achieved with an arrangement of a few Josephson junctions, known as superconducting nonlinear asymmetric inductive element (SNAIL) transmon~\cite{Frattini2017}, which is then sinusoidally driven at nearly twice the natural frequency of the oscillator. As a result, the system develops a double-well structure and a consequent twofold degenerate ground state that gives rise to the  cat states~\cite{Puri2017,Goto2018,Goto2019,Kwon2022,He2023,Grimm2020}.  A significant increase of the relaxation time has been achieved with this setup.

Driven nonlinear quantum oscillators have also been employed in theoretical studies of quantum activation~\cite{Marthaler2006,Lin2015}, quantum tunneling~\cite{Marthaler2007, Peano2012,Iyama2023,Prado2023,VenkatramanARXIV}, and photon-blockade phenomena~\cite{Roberts2020}. A better understanding of these systems can be achieved with the derivation of static effective Hamiltonians~\cite{DykmanBook2012,Venkatraman2022,XiaoXu2023}, as those used in the analysis of quantum tunneling~\cite{Prado2023,VenkatramanARXIV} and the coalescence of pairs of energy levels~\cite{Zhang2016} that result in excited state quantum phase transitions (``spectral kissing'')~\cite{Wang2020,Chavez2023,FrattiniPrep}.

Despite the various applications and the advances brought by Kerr parametric oscillators to quantum computation and quantum error correction~\cite{Campagne2020}, chaos can become a source of concern. The problem that the onset of chaos due to qubit-qubit interactions could cause to quantum computers was first raised in~\cite{Georgeot2000,Georgeot2000b,Braun2002,Silvestrov2001} and  reverberates in more recent studies about the scrambling of quantum information~\cite{Shenker2015,Maldacena2016JHEP,Garttner2017,Niknam2020}  
and the emergence of chaos in coupled Kerr parametric oscillators~\cite{Goto2021,Berke2022,Borner2023ARXIV}. 

Instead of interacting systems, our focus lies on the most basic element of the quantum computer, the qubit itself. The onset of global chaos in the high energy spectrum of transmon qubits was studied in~\cite{Cohen2023} and ways to suppress chaos in a general Hamiltonian for superconducting qubits was analyzed in~\cite{Burgelman2022}. Our concern is with the onset of local chaos that can develop at the core of the Kerr-cat qubit and melt it away. As the SNAIL transmons' nonlinearities are pushed to larger values~\cite{Blais2021,Frattini2017}, required for faster gates~\cite{Puri2017,Hillmann2020,Noguchi2020}, we show that the Kerr-cat qubit dangerously approaches its disintegration.

Driven nonlinear oscillators experimentally realized with the SNAIL transmon have so far been properly described by low-order static effective Hamiltonians. As the nonlinear effects increase, the static and driven pictures may still agree~\cite{GarciaPrep} if one considers higher orders terms in the expansion for the effective Hamiltonian~\cite{Venkatraman2022,xiao2023diagrammatic}, but this process eventually breaks down. When the drive and nonlinearities become sufficiently strong, chaos sets in and the oscillator can no longer be described by a time-independent Hamiltonian. 

If on the one hand, chaos puts limits on the Kerr-cat qubit,  on the other hand it opens a new direction of research for superconducting circuits~\cite{Neill2016}. Quantum chaos has received increasing attention in  fields that  range from quantum gravity and black holes to condensed matter and atomic physics due to its relationship with quantum dynamics, 
absence of localization, and thermalization.  Examples of quantum chaotic systems that have been experimentally realized include the kicked rotor~\cite{Moore1995}, the baker's map~\cite{Weinstein2002}, the kicked top~\cite{Chaudhury2009}, the kicked harmonic oscillator~\cite{Lemos2012}, and the driven pendulum~\cite{Hensinger2001,Steck2001}. Superconducting circuits offer unmatched advantages for 
the study of quantum chaos and its consequences, because both spectrum and dynamics can be measured simultaneously. The spectrum can be measured as a function of the control parameters, potentially allowing for the analysis of level statistics, and dynamics can be studied in phase space, which enables the evolution of out-of-time ordered correlators~\cite{Chavez2023} and Wigner functions.  Furthermore, the  classical limit is experimentally realizable.

In this work, we identify the border between regularity and chaos in the driven Kerr parametric oscillator, and estimate the parameter values for which the Kerr-cat qubit melts away. We also discuss how the qubit disintegration could be experimentally captured.
The analysis is based on the quasienergies and Floquet states of the driven nonlinear quantum oscillator implemented with the SNAIL transmon and is complemented with classical tools that include Poincar\'e sections and Lyapunov exponents.

\vskip 0.4 cm 


\large
{\raggedleft \bf{Quantum and Classical Hamiltonians}}
\normalsize
\\
The SNAIL transmon is an arrangement of Josephson junctions with a threaded magnetic flux that allows for tuning the nonlinearity of the system. Its Hamiltonian has a potential given by a sinusoidal function. As justified in Supplementary Note 1, to determine the onset of chaos in this system, it suffices to consider the Taylor expansion of the SNAIL potential up to fourth order, so the undriven part of the Hamiltonian is~\cite{FrattiniPrep,VenkatramanARXIV}
\begin{equation}
    \frac{\hat{H}_0}{\hbar} = \omega_0\hat{a}^\dagger\hat{a}+\frac{g_3}{3} (\hat{a}+\hat{a}^\dagger)^3 +
    \frac{g_4}{4} (\hat{a}+\hat{a}^\dagger)^4 .
\label{eq:H0}    
\end{equation}
In the equations above, $\omega_0$ is the bare frequency of the oscillator, $\hat{a}^\dagger$ and $\hat{a}$ are the bosonic creation and annihilation operators, and $g_{3},g_{4}\ll \omega_0$ are the coefficients of the third and fourth-rank nonlinearities \cite{FrattiniPrep,VenkatramanARXIV}. To create the Kerr-cat qubit, the system is periodically driven, so the total Hamiltonian is given by~\cite{FrattiniPrep,VenkatramanARXIV}
\begin{equation}
\frac{\hat{H}(t)}{\hbar} = \frac{\hat{H}_0}{\hbar} -i\Omega_d(\hat{a}-\hat{a}^\dagger)\cos \omega_d t ,
\label{eq:H}
\end{equation}
where  $\Omega_d$ is the amplitude of the sinusoidal drive, and $\omega_d$ is the driving frequency. We set $\hbar=1$.

The effective nonlinearity of the system, $K$, is determined by half the difference between the frequencies of the lowest energies of the undriven Hamiltonian, that is,
\begin{equation}
    K=(\omega_{1,0}-\omega_{2,1})/2,
    \label{eq:K}
\end{equation}
where $\omega_{i,j}=(E^{(0)}_i-E^{(0)}_j)$  and $E^{(0)}_i$ are the eigenvalues of $\hat{H}_0$. In the analysis below, we refer to $K$ as the Kerr nonlinearity and choose the control parameters $g_3$ and $g_4$ within ranges that are experimentally accessible. We stress that what we call $K$ here is an exact quantity, not the perturbative parameter used in effective Hamiltonians.

We use Floquet techniques~\cite{Holthaus2016} to analyze the periodically driven system in equation~(\ref{eq:H}). The Floquet operator over one period of the drive, $T_d=2\pi/\omega_d$, is denoted by 
\begin{equation}
\mathcal{U}  (T_d)|\mathcal{F}_j\rangle=\exp(-i\epsilon_j T_d)|\mathcal{F}_j\rangle,
\end{equation}
where $\epsilon_j$ are the quasienergies with $\epsilon_j T_d \in [- \pi,\pi]$   and $|\mathcal{F}_j\rangle$ are the 
Floquet states for $j\in[0,N-1]$, with $N$ being the truncated Hilbert space dimension.


The derivation of the classical limit of the quantum Hamiltonian in equation~(\ref{eq:H}) is shown in Methods. Using the canonical coordinates $(q,p)$, the Hamiltonian is written as
\begin{eqnarray}
h_{cl}(t)&=& h_0 + \sqrt{2}\Omega_{d} p \cos\left(\omega_dt\right),
 \label{eq:hcl}
\end{eqnarray}
where
\begin{equation}
h_0 =\frac{\omega_0}{2}\left(q^2 + p^2\right) +\frac{\sqrt{2^3}}{3}g_3q^3 +g_{4}q^4.
\label{eq:hcl0}
\end{equation}

\begin{figure*}[t!]
\bf{\centering
\includegraphics[scale=1.1]{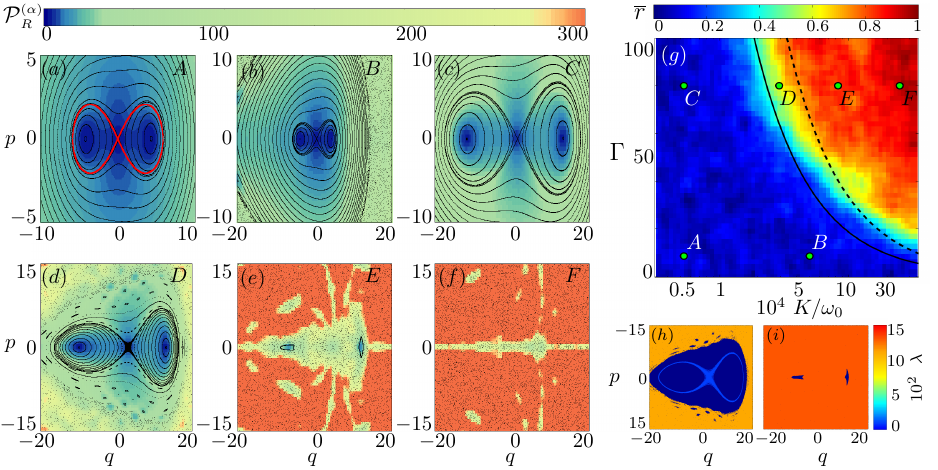}}
\caption{{\bf Regularity and chaos.} {\normalfont Quantum and classical analysis in phase space and the quasienergy spectrum for $\omega_d/\omega_0=1.999866$.  {\bf a - f}, Phase space analysis of the parameters indicated in Table~\ref{tab:points6}.  The black dots give the classical Poincar\'e sections for many different initial conditions, the red line in {\bf a} marks the separatrix that defines the Bernoulli lemniscate, colors from blue to orange indicate the values of the participation ratio of coherent states projected in the Floquet states. {\bf g}, Measure of quantum chaos given by the average ratio $\overline{r}$ of consecutive quasienergies spacings as a function of $K/\omega_0$ and $\Gamma$. The six points A-F marked in {\bf g}  are the same ones chosen for the phase spaces in {\bf a-f}. They were selected to illustrate the behavior in the regular, mixing, and chaotic regimes. The solid black curve in {\bf g} corresponds to equation~(\ref{eq:LEinside}) and indicates the parametric case, where the classical Lyapunov exponent becomes positive in the vicinity of the center of the lemniscate, while the black dashed line corresponds to equation~(\ref{eq:LEinout}) and indicates the parameters for which chaos sets in both inside and outside the original lemniscate, which by then has disappeared. {\bf h - i}, Lyapunov exponents for the same parameters used in (d)-(e). Zero Lyapunov exponent (dark blue) indicates regularity.}
}
\label{fig:1}
\end{figure*}

\vskip 0.4 cm  
\large
{\raggedleft \bf{Regularity to Chaos}} 
\normalsize
\\
We start our analysis by setting the frequency of the drive at nearly twice the natural frequency of the oscillator, $\omega_d\approx 2\omega_0$.  For this choice and the parameters used in the experiments~\cite{FrattiniPrep,VenkatramanARXIV}, the system can be described by a double-well metapotential, as illustrated in Fig.~\ref{fig:1}(a). The parameters are given in the first line of Table~\ref{tab:points6}, which defines the point A.

\begin{table}[h]
\begin{tabular}{@{}cccc@{}}
\toprule
 Point & $10^{4}K/\omega_0$ & $\Gamma$ & $n_{\text{min}}$\\
\hline
A & 0.53 & 8.5 & 8.079\\ 
B & 5.02 & 8.5 & 7.249 \\
C & 0.53 & 80 & 77.007 \\
D & 2.91 & 80 & 66.134 \\
E & 8.33 & 80 & 197.924 \\
F & 25 & 80 & 336.598  \\
\botrule
\end{tabular}
\caption{Parameters for the points A-F  marked in Fig.~\ref{fig:1}(g), whose phase diagrams are depicted in Figs.~\ref{fig:1}(a)-(f), and the corresponding values of $n_{\text{min}}$ obtained with equation~(\ref{eq:ngs}).}
\label{tab:points6}%
\end{table}

Black dots in Fig.~\ref{fig:1}(a) designate Poincar\'e sections. These points are obtained by evolving many different classical initial conditions according to equation~(\ref{eq:hcl}) and collecting the values of $q$ and $p$ at each time $T_d$. The curves that are formed with these points coincide with the energy contours of the classical limit of the static effective Hamiltonian investigated in~\cite{Chavez2023,FrattiniPrep,VenkatramanARXIV,GarciaPrep} [see equation~(\ref{eq:HKC}) in Methods]. The red curve in Fig.~\ref{fig:1}(a) is the Bernoulli lemniscate, which delineates the boundary of the double well and is characterized by the following two parameters: $\Pi=\Omega_d \omega_d/\left(\omega^2_d-\omega^2_0\right)$, where $\sqrt{2}\Pi$ is the distance from the center of the phase space to the center of the double well, and $\sqrt{2\Gamma}$, which is half the distance between the two minima of the wells, with $\Gamma= g_3 \Pi/K$. 
The symmetric ellipses within the lemniscate in Fig.~\ref{fig:1}(a) are centered at the minima of the metapotential\textcolor{blue}{,} at $(\pm q_{\text{min}} \!= \!\pm \sqrt{2\Gamma},\, p_{\text{min}}=0)$, and
the area within the lemniscate is equal to $4\Gamma$ (see Supplementary Note 2). Using the Bohr quantization rule and dimensionless coordinates $q$ and $p$, we thus have $\oint p dq = 2\pi n_{\text{in}}$, and the integer number of levels inside the lemniscate is given by~\cite{FrattiniPrep} 
\begin{equation}
n_{\text{in}}=2 \Gamma/\pi ,   
\label{eq:nb}
\end{equation}
which can be measured experimentally.

We color Fig.~\ref{fig:1}(a) according to the value of the participation ratio,
\begin{equation}
\mathcal{P}_{\mathcal{R}}^{(\alpha)}= \frac{1}{\sum_j \left| \bra{\alpha}\ket{\mathcal{F}_j}\right|^4} =  \frac{1}{\sum_{j} \left(\pi\mathcal{Q}_{\mathcal{F}_j}^{\alpha}\right)^2},
\label{Eq:PR}
\end{equation} 
for coherent states $|\alpha\rangle$ projected in the Floquet states, where $\hat{a}|\alpha\rangle=\alpha|\alpha\rangle$, with  $\alpha=(q+ip)/\sqrt{2}$, and $\mathcal{Q}_{\mathcal{F}_j}^{\alpha} = \left| \bra{\alpha}\ket{\mathcal{F}_j}\right|^2/\pi$ is the Husimi function of each Floquet state.  The participation ratio in equation~(\ref{Eq:PR}) measures the level of delocalization of a coherent state in the basis defined by $\ket{\mathcal{F}_j}$.   The most localized coherent states are those centered at the minima of the double-well metapotential, $|\pm\alpha_{\text{min}}\rangle$, and at its center   $(p,q)\simeq (0,0)$ \cite{Chavez2023}. They have the smallest values of $\mathcal{P}_R^{(\alpha)}$, which correspond to the darkest tones of blue in Fig.~\ref{fig:1}(a).

There are two quasidegenerate Floquet states, $|\mathcal{F}_{\text{min}}\rangle$, that are highly localized at the minima of the double wells and correspond to superpositions of the two opposite-phase coherent states,  $|\mathcal{F}_{\text{min}}\rangle\propto$ $|+\alpha_\text{min}\rangle \pm |-\alpha_\text{min}\rangle$ \cite{Cochrane1999,Mirrahimi2014}.  These states define the  Schr\"odinger cat states of the Kerr-cat qubit \cite{Grimm2020}. The expectation value of the number operator for these states is
\begin{equation}
    n_{\text{min}}=\langle \mathcal{F}_{\text{min}} |\hat{n}|\mathcal{F}_{\text{min}}\rangle \approx|\alpha_\text{min}|^2=\Gamma,
    \label{eq:ngs}
\end{equation}
which can be measured experimentally. This value is directly related with the number of states inside the lemniscate, $n_{\text{in}}$, given in equation~(\ref{eq:nb}). 


\vskip 0.4 cm  
{\raggedleft \bf{Kerr-cat qubit disintegration}}
\\
The portion of the phase space presented in Fig.~\ref{fig:1}(a) is characterized by periodic orbits, being therefore regular. However, a chaotic sea exists far away from the lemniscate, as shown in Supplementary Note 2. The analysis of global chaos would classify the system with the parameters of Fig.~\ref{fig:1}(a) as being in a mixed regime, but this is not our focus. We are concerned with local chaos, which can emerge around the phase space center and destroy the Kerr-cat qubit.  

To analyze the transition to chaos in the vicinity of the phase space center, we vary $\Gamma$ and $K/\omega_0$.  This is done so that the Kerr amplitude remains within  values that are experimentally accessible in the present or near future,  $K/\omega_0\in 33\times[ 10^{-6}, 10^{-4}]$
(see Methods). The parameter $\Gamma$ is varied by changing $\Pi$, while keeping $\omega_d \approx 2 \omega_0$.

To determine the onset of quantum chaos, we use the average ratio of consecutive quasienergy spacings~\cite{Oganesyan2007,Atas2013},
\begin{equation}
\tilde{r}=\frac{1}{N}\sum^N_j\text{min}\left(r_j,\frac{1}{r_j}\right), \hspace{0.2cm} \text{where}  \hspace{0.2cm}r_j=\frac{s_j}{s_{j-1}},
\end{equation}
and $s_j=\epsilon_{j+1}-\epsilon_j$. The spectra of chaotic systems are rigid and their levels are correlated, which results in Wigner-Dyson distributions for the spacings of neighboring levels. When the symmetries of the chaotic system comply with the circular orthogonal ensemble, $\tilde{r}_{\text{COE}}\approx 0.53$. Levels of regular systems are uncorrelated and follow Poisson statistics, $\tilde{r}_\text{P}\approx 0.39$. We compute the renormalized quantity, 
\begin{equation}
\bar{r}= \frac{\tilde{r}-\tilde{r}_\text{P}}{\tilde{r}_{\text{COE}}-\tilde{r}_\text{P}},
\label{eq:r}
\end{equation}
so that chaos entails $\bar{r}=1$ and regularity $\bar{r}=0$. 

In Fig.~\ref{fig:1}(g), we construct a map of regularity and chaos for the quantum system in equation~(\ref{eq:H}). The region in red indicates that $\bar{r} \approx 1$, so the system is chaotic. This region emerges for large values of the Kerr amplitude, $K/\omega_0$, and $\Gamma$. The region in blue indicates regularity.

The six points, A-F, marked in Fig.~\ref{fig:1}(g) are chosen for a more detailed analysis in Figs.~\ref{fig:1}(a)-(f) of their corresponding phase space structures (classical analysis) and of the level of delocalization of coherent states written in the basis of Floquet states (quantum analysis). Just as in Fig.~\ref{fig:1}(a), described above, the black dots in Figs.~\ref{fig:1}(b)-(f) are associated with the Poincar\'e sections and the colors give the values of the participation ratio of coherent states projected in the Floquet states.

Points A, B, and C are in the regular regime. The lemniscate in Fig.~\ref{fig:1}(a) persists in Figs.~\ref{fig:1}(b)-(c), although it becomes more asymmetric. Notice that the scales in Fig.~\ref{fig:1}(a) are not the same as in Figs.~\ref{fig:1}(b)-(c) .

Point B corresponds to a large value of the Kerr amplitude. As seen at the edges of Fig.~\ref{fig:1}(b), the periodic orbits disappear, giving space to black dots. In the classical limit, this region of phase space gives positive Lyapunov exponents, which implies chaos. In spite of that, the structure of the Kerr-cat qubit survives and the value of $n_{\text{min}}$ remains close to $\Gamma$, as seen in Table~\ref{tab:points6}. The resilience of the Kerr-cat qubit to a range of values of the the Kerr nonlinearity should be reassuring to the parametric quantum computation community (see also Supplementary Note 3). 

Point C shows what happens to point B as one approaches the classical limit, which is done by broadening the wells. By increasing  $\Gamma$ while keeping $\Gamma K/\omega_0$ constant, we enlarge the wells without changing their shape and increase the number of levels within (cf. the values of $n_{\text{min}}$ for B and C in Table~\ref{tab:points6}), thus approaching the classical picture.

Point D provides the main message of this work. The center of the double well, which is a hyperbolic point in Figs.~\ref{fig:1}(a)-(c), is no longer a single point in Fig.~\ref{fig:1}(d). Chaos now exists not only far away from the double-well structure, but right at the center of  the lemniscate, indicating the beginning of its disintegration. At this stage, any activation between wells~\cite{Marthaler2006,FrattiniPrep} will happen through the chaotic region. We now have chaos and islands of stability around the structure of the asymmetric double well and chaos at its center.

To make the onset of local chaos in Fig.~\ref{fig:1}(d) even more evident, we replicate this panel in Fig.~\ref{fig:1}(h), but now color it with the values of the Lyapunov exponent, $\lambda$, obtained with the classical system in equation~(\ref{eq:hcl}). There are two distinct Lyapunov exponents, the one beyond the double well, associated with global chaos, and the one right at its center, responsible for melting the qubit away.

\begin{figure*}[t]
\bf{
\centering
    \includegraphics[scale=1.1]{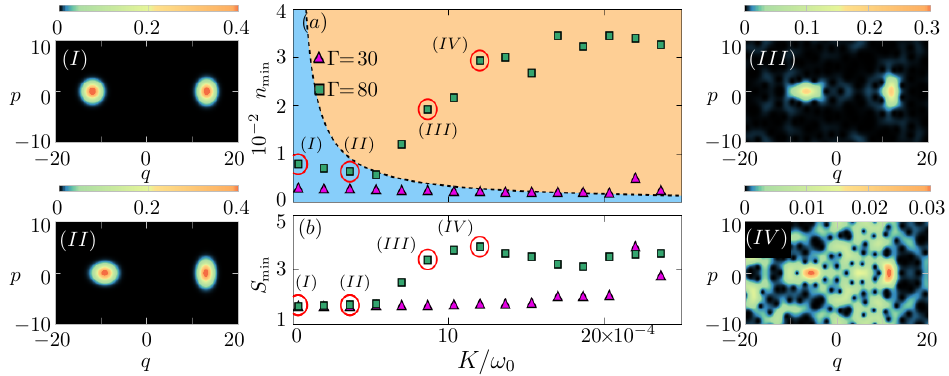}}
	\caption{{\bf Kerr-cat qubit disintegration.} {\normalfont {\bf a}, Expectation value of the number operator, $n_{\text{min}}$, and {\bf b}, Shannon entropy, $S_{\text{min}}$, for the Floquet state $|\mathcal{F}_{\text{min}}\rangle$. The two quantities are shown as a function of the Kerr amplitude $K/\omega_0$ for $\Gamma=30$ (triangles) and $\Gamma=80$ (squares). In {\bf a}: The blue background indicates regular region and the orange background indicates chaotic region; they are separated by the same black dashed line shown in  Fig. \ref{fig:1}(g).
     Panels (I), (II), (III), and (IV) depict the Husimi functions for the Floquet state $|\mathcal{F}_{\text{min}}\rangle$ indicated in {\bf a} and {\bf b} as points  (I), (II), (III), and (IV)  with $K/\omega_0=\{0.33,3.66,8.66, 12\} \times 10^{-4}$. }
	}
		\label{fig:2}
	\end{figure*}

The values of $\Gamma$ and $K/\omega_0$ beyond which the intermediate regime between regularity and local chaos emerges follows the black solid line in Fig.~\ref{fig:1}(g) given by
\begin{equation}
\Gamma K/\omega_0 = \frac{g_3 \Omega_d \omega_d}{\omega_0 (\omega_d^2 - \omega_0^2) } \simeq 0.0187. 
\label{eq:LEinside}
\end{equation}
This line marks the parameters' values, where the Lyapunov exponent first gets positive in the vicinity of the phase-space center, remaining separated from the region of global chaos. This is a theoretical line, that can be refined according to the particularities of each experimental setup. It indicates that, despite the transition to chaos, there is still ample space for the stabilization of Schr\"odinger cat states and for reaching large values of $K$, which are needed for fast gates.

By increasing the strength of the nonlinearities and drive, the chaotic sea, which was once far away from the double well, expands and eventually merges together with the chaotic region of the phase-space center. This is illustrated with points E-F in Figs.~\ref{fig:1}(e)-(f).  In Fig.~\ref{fig:1}(e), two small islands of regularity  reminiscent of the double well persist. They are also visible in the corresponding Fig.~\ref{fig:1}(i), which depicts the Lyapunov exponent for the parameters of point E. In contrast to  Fig.~\ref{fig:1}(i), there is now a single positive Lyapunov exponent.
In Fig.~\ref{fig:1}(f), the values of $\mathcal{P}_{\mathcal{R}}^{(\alpha)}$  indicate near ergodicity.

In Fig.~\ref{fig:1}(g), we draw a dashed black line to indicate the parameters' values for which chaos close to the phase-space center and around the double well merge together. Similarly to equation~(\ref{eq:LEinside}), the analysis is based on the values of the Lyapunov exponents and the equation for the dashed line is given by 
\begin{equation}
\Gamma K/\omega_0 = \frac{g_3 \Omega_d \omega_d}{\omega_0 (\omega_d^2 - \omega_0^2) } \simeq 0.03347.
    \label{eq:LEinout}
\end{equation}
We notice, however, that the Taylor expansion used in equation~(\ref{eq:H0}) works well for values of the nonlinearities and drive up to equation~(\ref{eq:LEinside}). Beyond that, the full sinusoidal SNAIL potential should offer a better picture of the chaotic system. 

The analysis in Fig.~\ref{fig:1} was performed using a relation between $g_3$ and $g_4$ that ensures that the parameters in Fig.~\ref{fig:1}(a) reproduce the physics in~\cite{FrattiniPrep}, where the second-order static effective Hamiltonian describes very well the experiment. There are numerous other possibilities for varying the parameters, many within experimental capabilities. Nevertheless, as discussed in Supplementary Note 3, they should lead to results comparable to those in Fig.~\ref{fig:1}. The transition to chaos is unavoidable, although one may be able to shift the values for the threshold between regularity and chaos in equation~(\ref{eq:LEinside}).

\vskip 0.4 cm  
\large
{\raggedleft \bf{Chaos Detection}}
\normalsize
\\
The experiment with the superconducting circuit performed in~\cite{FrattiniPrep} measured the energy levels of the driven nonlinear oscillator as a function of the control parameter. However, the number of levels currently accessible to the experiment is not sufficient for the analysis of level statistics, as done in Fig.~\ref{fig:1}(g). To circumvent this issue, we propose a way to  detect the transition to chaos that avoids the analysis of the quasienergy spectrum and focuses instead on the properties of the Floquet state $|\mathcal{F}_{\text{min}}\rangle$. When the system is in the regular regime, this state coincides with the Schr\"odinger cat state and is highly localized at the minima of the wells. As the nonlinearities increase and $|\mathcal{F}_{\text{min}}\rangle$ spreads in phase space, we can be sure that chaos has already set in.

In Fig.~\ref{fig:2}(a), we show $n_{\text{min}}=\langle \mathcal{F}_{\text{min}} |\hat{n}|\mathcal{F}_{\text{min}}\rangle $ as a function of $K/\omega_0$ for  $\Gamma=30$ (triangles) and $\Gamma=80$ (squares). In the presence of the double well, $n_{\text{min}} \sim \Gamma$, as given by equation~(\ref{eq:ngs}). The background of the figure is colored according to the results in Fig.~\ref{fig:1}(g), so the region in blue is regular and orange one indicates chaos. The dashed black line separating the two regions is the same as in Fig.~\ref{fig:1}(g).
The analysis is complemented with Fig.~\ref{fig:2}(b), which shows the behavior of the Shannon entropy for the Floquet state $|\mathcal{F}_{\text{min}}\rangle$ projected in the coherent states, 
\begin{equation} 
S_{\text{min}} = - \frac{1}{\pi}\int \mathcal{Q}^\alpha_{{\mathcal{F}_{\text{min}}}} \ln \left( \mathcal{Q}^\alpha_{{\mathcal{F}_{\text{min}}}}\right)\, d^2\alpha ,
\end{equation}
as function of $K/\omega_0$ for $\Gamma=30$ (triangles) and $\Gamma=80$ (squares).

We start by describing the results for $\Gamma=80$ (squares) in Fig.~\ref{fig:2}(a). In the regular regime, $n_{\text{min}}$  decays linearly with the Kerr amplitude. To better explain this behavior, we select points (I) and (II) and depict their respective Husimi functions on the left panels (I) and (II). As expected, the Husimi functions for these two Floquet states $|\mathcal{F}_{\text{min}}\rangle$ are localized at the minima of the double well, at $q=\pm \sqrt{2\Gamma}\approx \pm 13$. Comparing panel (I) and panel (II), we see that as $K/\omega_0$ increases, the structure of the Husimi function becomes more asymmetric and the area of the lemniscate decreases, which reduces the value of $n_{\text{min}}$. Since the Husimi functions remain localized in panels (I) and (II), the values of the Shannon entropy for these two cases in Fig.~\ref{fig:2}(b) are comparable.

As we enter the chaotic region for $\Gamma=80$, $n_{\text{min}}$ in  Fig.~\ref{fig:2}(a) and $S_{\text{min}}$ in Fig.~\ref{fig:2}(b)  grow with $K/\omega_0$.
 This can be understood from the Husimi functions for the points (III) and (IV), shown on the panels to the right of Figs.~\ref{fig:2}(a)-(b). The parameters for point (III) are equivalent to those in Fig.~\ref{fig:1}(e), where there are two islands of instability close to the original minima of the double well. This explains why $|\mathcal{F}_{\text{min}}\rangle$  in panel (III)  shows some level of confinement around the islands,  although the state is visibly more delocalized than those in panels (I) and (II). The parameters for point (IV) are equivalent to those in Fig.~\ref{fig:1}(f), where the system approaches ergodicity, so the Husimi function in panel (IV) is spread out. 
 

The behavior of $n_{\text{min}}$ and $S_{\text{min}}$ as a function of the Kerr amplitude for $\Gamma=30$ (triangles) in Figs.~\ref{fig:2}(a)-(b) is similar to that for $\Gamma=80$. The difference lies in the values of the Kerr amplitude required for the onset of chaos and the consequent growth of $n_{\text{min}}$ and $S_{\text{min}}$, which are larger than  for $\Gamma=80$.

The disintegration of the double well  can then be detected from the analysis of the spread of the Schr\"odinger cat states in phase space. This can be done by directly  investigating the Husimi or Wigner functions of these states in phase space for different values of the system parameters, or by quantifying their spread with the occupation number $n_{\text{min}}$ or an entropy, such as  $S_{\text{min}}$. The growth of $n_{\text{min}}$ and  $S_{\text{min}}$ signals the system's departure from the regular to the chaotic regime.

\vskip 0.4 cm  
\large
\noindent {\bf Conclusion} 
\\
\normalsize
Our work brings to light the risk posed by the onset of chaos for Kerr parametric oscillators, which puts a limit on the ranges of parameters that can be employed for qubit implementation. Combining quantum and classical analysis, we determined the threshold for the rupture of the Kerr-cat qubit, which happens when chaos first sets in around the center of the qubit double-well structure.  Important extensions to this work include the role of dissipation~\cite{Habib1998,Ferrari2023ARXIV} and the analysis of the limitations that chaos may impose to parametric gates in transmon and fluxonium arrays. 

By increasing the nonlinearities and driving amplitude, we showed that the Schr\"odinger cat states of the Kerr-cat qubit, which are initially confined at the bottom of the wells, spread and eventually  disintegrate. Once these states are lost, chaos is certain to have spread throughout the phase space. The process of disintegration could be experimentally observed with the currently available technology by measuring the Wigner functions of the cat states.  

The results in this work indicate that the platform of superconducting circuits allows either to engineer bosonic qubits for quantum technologies or to induce chaos to address fundamental questions. This opens up a new avenue of research for superconducting circuits. They could be used, for example, to investigate how chaos affects the spread of quantum information in phase space and whether chaos can enhance the tunneling rate between islands of stability.

\vskip 0.8 cm  
\noindent {\bf METHODS} 

\noindent {\bf Quantum and Classical Hamiltonian}

To derive the classical Hamiltonian, we write \begin{equation}
\label{eq:a_z}
\hat{a}= \frac{1}{\sqrt{2 \hbar_\text{eff}} }\left(\hat{q}+i  \hat{p}\right),
\end{equation}
and
$$
[\hat{q},\hat{p}]=i\hbar_\text{eff},
$$
where $\hbar_\text{eff}$ depends on experimental parameters as explained in the Supplementary Note 1. The classical limit is reached by taking $\hbar_\text{eff} \rightarrow 0$, since $\hat q\rightarrow q$ and $\hat p\rightarrow p$. This way, the quantum Hamiltonian,
\begin{eqnarray}
\frac{\hat{H}(t)}{\hbar } &=&  \frac{\omega_0}{2\hbar_\text{eff}} \left(\hat{q}-i  \hat{p}\right) \left(\hat{q}+i  \hat{p}\right) + \sum^{4}_{m=3}\frac{g_m}{m}\left( \sqrt{\frac{2}{\hbar_\text{eff}}}\hat{q}\right)^m\nonumber \\
 &+& \Omega_d  \frac{\sqrt{2}}{\hbar_\text{eff}}\,\hat{p}\,\cos \omega_d t,
\end{eqnarray}
leads to the classical Hamiltonian,
\begin{eqnarray}
h_{cl}(t)&=& \frac{\omega_{0}^{cl}}{2}\left(q^2 + p^2\right) +\frac{2\sqrt{2}}{3}g_{3}^{cl}q^3 +g_{4}^{cl}q^4  \nonumber \\
 &+&  \sqrt{2}\Omega_{d}^{cl} p \cos\left(\omega_dt\right),
 \label{eq:hclA}
\end{eqnarray}
where 
$$\omega_0=\omega_{0}^{cl}\hbar_\text{eff}, \hspace{0.2 cm} g_3=g_{3}^{cl}\sqrt{\hbar_\text{eff}}, $$
$$
g_4=g_{4}^{cl}\hbar_\text{eff}^2, \hspace{0.3 cm} \text{and} \hspace{0.3 cm} \Omega_d = \Omega_{d}^{cl}\sqrt{\hbar_\text{eff}}.$$ 
The classical static Hamiltonian
\begin{equation}
h_0 =\frac{\omega_0}{2}\left(q^2 + p^2\right) +\frac{\sqrt{2^3}}{3}g_3q^3 +g_{4}q^4.
\label{eq:SMh0}
\end{equation}
describes a quartic asymmetric oscillator, that presents three stationary (critical) points (more details in Supplementary Note 1).

\vskip 0.3 cm
\noindent {\bf Control parameters}

The values of $K/\omega_0$ are varied parametrically  by changing $g_3/\omega_0$ and $g_4/\omega_0$  according to the equation
\begin{equation}
g_4 = \frac{20 g_3^2}{69 \omega_0}.
\label{eq:g4g3}
\end{equation}
This choice is made to guarantee that we reproduce the scenario in~\cite{FrattiniPrep}, where the second-order effective Hamiltonian describes very well the experiment. The second-order effective Hamiltonian is given by~\cite{FrattiniPrep},
\begin{equation}
\label{eq:HKC} \frac{\hat{H}^{(2)}_{\mathrm{eff}}}{\hbar} = - K^{(2)} \hat a^{\dagger2} \hat a^2 +\epsilon_2^{(2)} (\hat a^{\dagger2} + \hat a^{2}) ,
\end{equation}
where 
\begin{equation}
K^{(2)}= -\frac{3g_4}{2} +  \frac{10g_3^2}{3\omega_0} ,
\label{eq:K2}
\end{equation}
and $\epsilon_2^{(2)} = 2 g_3 \Omega_d/(3\omega_0)$. 
Equation~(\ref{eq:g4g3}) is the same as equation~(\ref{eq:K2}) when $K^{(2)}= 10  g_4$. In  Supplementary Note 3, we show what happens to the analysis in Fig.~\ref{fig:1}(g) for other choices of ${\cal C}$ in $K^{(2)}= {\cal C} \, g_4$.

\vskip 0.3 cm 
\noindent  {\bf ACKNOWLEDGMENTS}

\noindent
We thank Roman Baskov, Rodrigo Corti\~nas, and Michel Devoret for valuable suggestions and comments on the manuscript.
This research was supported by the
NSF CCI grant (Award Number 2124511). D.A.W and I.G.-M. received support from CONICET (Grant
No. PIP 11220200100568CO), UBACyT (Grant
No. 20020170100234BA) and ANCyPT (Grants
No. PICT-2020-SERIEA-00740 and PICT-2020-
SERIEA-01082). F.P.B. thanks Grant PID2022-136228NB-C21 funded by MCIN/AEI/
10.13039/501100011033 and, as appropriate, by “ERDF A way of making Europe”, by
the “European Union” or by the “European Union NextGenerationEU/PRTR”. Computing resources supporting this work were partially provided
by the CEAFMC and Universidad de Huelva High Performance Computer (HPC@UHU) located in the 
Campus Universitario el Carmen and funded by FEDER/MINECO Project UNHU-15CE-2848.

\vskip 0.3 cm


%


\newpage

\onecolumngrid
\vspace*{0.4cm}
\begin{center}

{\large \bf Supplementary Information: Driving superconducting qubits into chaos}
\vspace{0.6cm}

Jorge Ch\'avez-Carlos$^1$, Miguel A. Prado Reynoso$^1$, Ignacio Garc\'ia-Mata$^2$, Victor
S. Batista$^3$, Francisco P\'erez-Bernal$^4$, Diego A. Wisniacki$^5$, and Lea F. Santos$^6$

$^1${\it Department of Physics, University of Connecticut, Storrs, Connecticut 06269, USA}

$^2${\it Instituto de Investigaciones Físicas de Mar del Plata (IFIMAR), Facultad de Ciencias Exactas y Naturales,
Universidad Nacional de Mar del Plata, CONICET, 7600 Mar del Plata, Argentina}

$^3${\it Department of Chemistry, Yale University, 
P.O. Box 208107, New Haven, Connecticut 06520-8107, USA}

$^4${\it Departamento de Ciencias Integradas y Centro de Estudios Avanzados en F\'isica, Matem\'aticas y Computaci\'on, Universidad de Huelva, Huelva 21071, Spain}

$^5${\it Departamento de F\'isica ``J. J. Giambiagi'' and IFIBA, FCEyN,
Universidad de Buenos Aires, 1428 Buenos Aires, Argentina}

\end{center}

\vspace{0.6cm}

\section{Driven SNAIL transmon}
\label{app:snail}

The Hamiltonian of the experimental driven circuit with $M$ SNAILs is given by
\begin{equation}
\frac{\hat{H}(t)}{\hbar} =
4E_C\hat{n}^2
+U_M\left(\hat{\varphi}\right)+\sqrt{2}\tilde{\Omega}_d \hat{n}\cos(\omega_d t) .
\label{eq:snail}
\end{equation}
The inductive energy of a single SNAIL is\begin{equation}
U_\text{SNAIL}(\hat{\varphi}_s)=
E_J\left[-\alpha \cos(\hat{\varphi}_s) - 
	m\cos\left(\frac{\varphi_\text{ext}-\hat{\varphi}_s}{m}\right)
\right] ,
\label{eq:Usnail}
\end{equation}
$E_C=e^2/2C$ is the Coulomb charging energy of the junction with capacitance $C$, $e$ is the electron charge, $E_J$ is the Josephson energy, $\varphi_\text{ext}=2\pi \Phi_\text{ext}/\Phi_0$ is the reduced applied magnetic flux, $\Phi_0=h/2e$ is the magnetic flux quantum, $\hat{\varphi}_s$ is the phase drop across the single SNAIL, and the drive is defined by its amplitude $\tilde{\Omega}$ and its frequency $\omega_d$.
The operators $\hat{n}$ and $\hat{\varphi}$ describe the reduced charge on the capacitance and its conjugate, the reduced flux operator, where $[\hat{\varphi},\hat{n}]=i$.
The single SNAIL consists of a superconducting loop of $m$ large Josephson junctions and a single smaller junction (tunneling energies $E_J$ and $\alpha E_J$, respectively), which is threaded with a DC magnetic flux $\Phi_\text{ext}$. 
For an array of $M$ SNAILs, the effective potential reads as
\begin{equation}  U_M(\hat{\varphi})=MU_\text{SNAIL}(\hat{\varphi}_s[\hat{\varphi}])+\dfrac{1}{2}E_L\left(\hat{\varphi} -M\hat{\varphi}_s[\hat{\varphi}] \right)^2 ,
\label{eq:MUsnail}
\end{equation}
where $E_L$ is the energy of the linear inductance (or stray inductance) in the system and $\varphi_s[\varphi]$ is defined by the equation
\begin{equation}
    \alpha \sin \varphi_s-\sin\left(\frac{\varphi_\text{ext}-\varphi_s}{m}\right)+\xi_J(M\varphi_s-\varphi)=0 ,
\end{equation}
where $\xi_J=L_J/L$ is the ratio between the inductance of the big junction in the SNAIL and the linear inductance $L$.

\subsection{Taylor expansion of the SNAIL potential}

By Taylor expanding the potential of the single SNAIL in equation~(\ref{eq:Usnail}) around  the minimum $\varphi_s[\bar{\varphi}_\text{min}]=\varphi_\text{min}$, we have
\begin{equation}
U_\text{SNAIL}\left(\hat{\varphi}+\varphi_\text{min}\right)\approx E_J\left(\frac{c_2}{2}\hat{\varphi}^2
+\frac{c_3}{3!}\hat{\varphi}^3
+\frac{c_2}{4!}\hat{\varphi}^4 \cdots\right),
\qquad\qquad
c_n=\frac{\partial^n U\left(\varphi_\text{min}\right)}{\partial \varphi^n} ,
\end{equation}
where
\begin{equation}
\begin{split}
c_0 &=-\alpha\cos\left(\varphi_\text{min}\right)
-m\cos\left(\frac{\varphi_\text{ext} - \varphi_\text{min}}{m}\right), \\
c_1 &=\alpha\sin\left(\varphi_\text{min}\right)
- \sin\left(\frac{\varphi_\text{ext} - \varphi_\text{min}}{m}\right), \\
c_2 &=\alpha\cos\left(\varphi_\text{min}\right)
+ \frac{1}{m}\cos\left(\frac{\varphi_\text{ext} - \varphi_\text{min}}{m}\right),
\end{split} \qquad\qquad
\begin{split}
c_3 &=-\alpha\sin\left(\varphi_\text{min}\right)
+ \frac{1}{m^2}\sin\left(\frac{\varphi_\text{ext} - \varphi_\text{min}}{m}\right), \\
c_4 &=-\alpha\cos\left(\varphi_\text{min}\right)
- \frac{1}{m^3}\cos\left(\frac{\varphi_\text{ext} - \varphi_\text{min}}{m}\right) ,
\end{split}
\end{equation}
and $\varphi_\text{min}$ obeys the transcendental equation
\begin{equation}
c_1=\alpha \sin \left(\varphi_\text{min}\right) - \sin\left(\frac{\varphi_\text{ext} - \varphi_\text{min}}{m}\right)=0 .
\end{equation}

For an array of $M$ SNAILs, the potential energy of the SNAIL in equation (\ref{eq:MUsnail}) can also be expanded, leading to the following time-independent part of the total Hamiltonian in equation~(\ref{eq:snail}),
\begin{equation}
\frac{\hat{H}_0}{\hbar} =
4E_C\hat{n}^2
+E_J\left(\frac{\bar{c}_2}{2!}\hat{\varphi}^2+\frac{\bar{c}_3}{3!}\hat{\varphi}^3+\frac{\bar{c}_4}{4!}\hat{\varphi}^4+\cdots\right),
\label{eq:snailexp}
\end{equation}
where the coefficients $\bar{c}_n$ can be replaced  with the coefficients $c_n$ introduced in \cite{Frattini2018} as
\begin{equation}
\begin{split}
\bar{c}_2 &=\frac{p}{M}c_2, \\
\bar{c}_3 &=\frac{p^3}{M^2}c_3, \\
\bar{c}_4 &=\frac{p^4}{M^3}\left[c_4-\frac{3c_3^2}{c_2}(1-p)\right],
\end{split}
\end{equation}
where $p=\frac{M\xi_J}{c_2+M\xi_J}$.
Only the third and fourth-rank nonlinearities were relevant in the experiments in~\cite{Frattini2018,Sivak2019,FrattiniPrep,VenkatramanARXIV}, as also discussed theoretically  in~\cite{Hillmann2020}. We show below the relevance of this expansion over the local dynamics in phase space.
 
Introducing the dimensionless operators
\begin{equation}
\hat{X}=\frac{\hat{\varphi}}{\sqrt{2}\hbar_\text{eff}},
\qquad
\hat{P}=\sqrt{2}\hbar_\text{eff}\,\hat{n} ,
\end{equation}
 with $[\hat{X},\hat{P}]=i$, and truncating equation~(\ref{eq:snailexp}) at fourth order in $\hat{\varphi}$, we arrive at the following total time-dependent Hamiltonian
\begin{equation}
\frac{\hat{H}(t)}{\hbar }=\omega_0 \left(  \frac{\hat{P}^2}{2} + \frac{\hat{X}^2}{2}\right) 
+\frac{2\sqrt{2}g_3}{3}\hat{X}^3
+g_4\hat{X}^4
+\sqrt{2}\Omega_d \cos\left(\omega_d t\right)\hat{P} ,
\label{eq:SM_classical}
\end{equation}
where 
\begin{equation}
\omega_0=\sqrt{8\bar{c}_2E_C E_J }=2\hbar_\text{eff}^2\bar{c}_2E_J,\qquad g_3=\frac{\hbar_\text{eff}^3\bar{c}_3E_J}{2},
\qquad
g_4=\frac{\hbar_\text{eff}^4\bar{c}_4E_J}{6},
\qquad
\Omega_d=\frac{\tilde{\Omega}_d}{\sqrt{2}\hbar_\text{eff}},
\end{equation}
and 
\begin{equation}
\hbar_\text{eff}=\left(\frac{2E_C}{\bar{c}_2E_J}\right)^{1/4} .
\end{equation}

\begin{figure}[h]
   \bf{\centering
\includegraphics[scale=1.0]{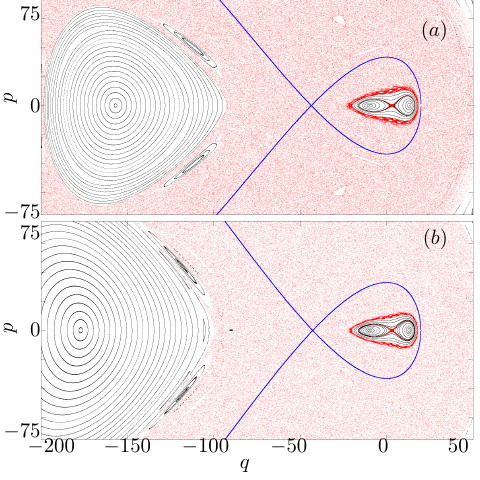}}
    \caption{{\bf Comparison between full and expanded SNAIL potential in phase space.}  {\bf a}{\normalfont,  Poincar\'e sections from the classical Hamiltonian with the full SNAIL potential. {\bf b}, Poincar\'e sections from classical Hamiltonian with the SNAIL potential expanded to fourth order. Both plots have the parameters corresponding to the point D in the Table 1 of the main text. Black points are used for regular orbits and red points are used for chaotic orbits. The blue solid line represents the separatrix for the non-expanded potential ({\bf a}) and the potential expanded to fourth order ({\bf b}).}
    }
    \label{fig:SM01} 
\end{figure}
Using the classical limit described in Methods, we compare the Poincar\'e section obtained for the full potential [Fig.~S~\ref{fig:SM01}(a)] with the Poincar\'e section obtained for the Taylor expanded potential [Fig.~S~\ref{fig:SM01}(b)] for the parameters used in the case of point D in Fig.~1 of the main text. This is the point at the transition region between integrability and chaos, which is the focus of this work. We see that the results in the region of the double well structure are equivalent. This means that for the analysis of the transition to chaos, it suffices to consider the Taylor expanded potential as done in the main text.

\subsection{Phase-space volume rescaling}

Using the following relation
\begin{equation}
\hat{X}=\hat{q}/\sqrt{\hbar_\text{eff}}, \qquad
\hat{P}=\hat{p}/\sqrt{\hbar_\text{eff}} 
\end{equation}
in $[\hat{X},\hat{P}]=i$, we get $[\hat{q},\hat{p}]=i\hbar_\text{eff}$. The quantum Hamiltonian in terms of $\hat{q}$ and $\hat{p}$ becomes
\begin{equation}
\begin{split}
\frac{\hat{H}(t)}{\hbar}=\frac{\omega_0}{\hbar_\text{eff}}\left(\frac{\hat{q}^2 + \hat{p}^2}{2}\right)
+\frac{2\sqrt{2} g_3}{3\sqrt{\hbar_\text{eff}^3}}\hat{q}^3
+  \frac{g_4}{\hbar_\text{eff}^2}\hat{q}^4
+
\sqrt{\frac{2}{\hbar_\text{eff}}}\Omega_d\hat{p}\cos\left(\omega_d t\right).
\end{split}
\end{equation}
By decreasing $\hbar_\text{eff}$, the double-well structure grows and the number of states within increases, thus bringing the system closer to the classical limit. An example of this scenario is given in Fig.~1 of the main text, as we move from point B to point C [see Fig.~1(b) and Fig.1~(c) of the main text].

Defining $\hat{a}=\dfrac{1}{\sqrt{2\hbar_\text{eff}}}(\hat{q}+i\hat{p})$, we can write the Hamiltonian as
\begin{equation}
\frac{\hat{H}(t)}{\hbar}=\omega_0 \hat{a}^\dagger\hat{a}+\frac{g_3}{3}(\hat{a}+\hat{a}^\dagger)^3
+ \frac{g_4}{4}(\hat{a}+\hat{a}^\dagger)^4
-i\Omega_d(\hat{a}-\hat{a}^\dagger)\cos\left(\omega_d t\right).
\end{equation}
From the parameters $\omega_0$, $g_3$, $g_4$ and $\Omega_d$, it is possible to get the classical limit using
\begin{equation}
\omega_0=\omega_0^{cl}\hbar_\text{eff},\qquad
g_3=\sqrt{\hbar_\text{eff}^3}g_3^{cl},\qquad
g_4=\hbar_\text{eff}^2g_4^{cl},\qquad
\Omega_d=\sqrt{\hbar_\text{eff}}\Omega_d^{cl} ,
\end{equation}
where the classical parameters have the superscript $cl$.

\section{Emergence of the Bernoulli lemniscate}
\label{app:BL}

To better understand the origin of the lemniscate in Fig. 1(a) of the main text and where it emerges in phase space, let us start by analyzing the classical static Hamiltonian in  equation (6) of the main text,
\begin{equation}
h_0 =\frac{\omega_0}{2}\left(q^2 + p^2\right) +\frac{\sqrt{2^3}}{3}g_3q^3 +g_{4}q^4.
\label{eq:SMh0}
\end{equation}
This Hamiltonian describes a quartic asymmetric oscillator that presents three stationary (critical) points with $p=0$. They are the minima 
\begin{eqnarray}
(q_0,p_0)&=&(0,0) ,\nonumber \\
(q_1,p_1)&=&(d_{-},0), \nonumber
\end{eqnarray}
and the hyperbolic point 
\[
(q_2,p_2)=(d_{+},0),
\]where $d_\pm=\sqrt{2}\left(-g_3\pm\sqrt{g^2_3-2g_4\omega_0}\right)/(4g_4)$. The condition $g_3^2-2g_4\omega_0>0$ ensures that  $d_\pm$ is real.

The linearized Hamilton equations around a critical point  $\{q_c, p_c\}$ of $h_0$ satisfies the following linear differential equations,
\begin{eqnarray}
\begin{pmatrix}
 \dot{q} \\
\dot{p}
\end{pmatrix}
=   \begin{pmatrix}
0 & \omega_0 \\
-\omega_0\!-\!4\sqrt{2}g_3q_c\!-\!12g_4q_c^2 &   0 \\
\end{pmatrix}
\hspace{-0.1 cm }
\begin{pmatrix}
q\!-\!q_c \\
p\!-\!q_c
\end{pmatrix}. 
\label{eq:linear}
\end{eqnarray}
The stability or instability around  $\{q_c, p_c\}$ is determined by the eigenvalues $\lambda_m$ of the matrix constructed in equation~(\ref{eq:linear}). If the eigenvalues are complex numbers, $\lambda_m=i\,\Tilde{\omega}_m$, the orbits in the neighborhood of the critical point are periodic and have frequencies $\Tilde{\omega}_m$.
 If the eigenvalues of the matrix are real, then the critical point is unstable and its Lyapunov exponent is equal to $\text{max}(\lambda_m)$.

\begin{figure}[h]
    \bf{\centering  \includegraphics[scale=1.0]{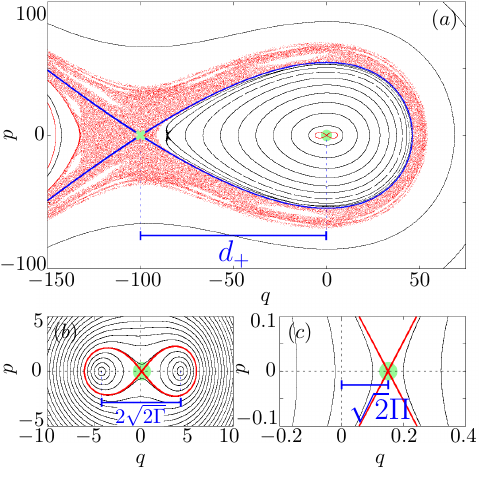} }
    \caption{{\bf Emergence of the Kerr cat qubit.}  {\bf a}{\normalfont,  Phase space metapotential of the classical Hamiltonian $h_{cl}(t)$ in equation~(\ref{eq:SMh0}) representing a large asymmetric double well. Black points are used for regular orbits. The red points indicate orbits with positive Lyapunov exponents (chaos). The two green symbols indicate the critical points: circle for $(q_0,p_0)=(0,0)$, and cross for $(q_2, p_2)=(d_{+},0)$. The blue line is the separatrix of the asymmetric double well.  {\bf b}, Enlarged image of panel  {\bf a} close to the point $(0,0)$, providing a view of the additional symmetric double well that emerges at the phase space center.  The red line is the Bernoulli lemniscate. The distance between the two minima is $2 \sqrt{2 \Gamma}$.  {\bf c}, Enlarged image of panel  {\bf b} close to the point $(0,0)$. The distance between the phase space center $(0,0)$ and the hyperbolic point of the Bernoulli lemniscate is $\sqrt{2} \Pi$.}
  }
    \label{fig:SM02} 
\end{figure}

In Fig.~S~\ref{fig:SM02}(a), we show the Poincar\'e sections (black lines)  for the driven system described by $h_{cl}(t)$ in equation (\ref{eq:SMh0}) with a frequency $\omega_d$ that is nearly twice $\omega_0$ and with the parameters used in the experiment in~\cite{FrattiniPrep} and in Fig.~1(a) of the main text. The stationary points of $h_0$, $(q_0,p_0)=(0,0)$ and $(q_2,p_2)=(d_{+},0)$ are marked with green symbols: circle for $(0,0)$ and cross for $(q_2, p_2)$.  The blue line crossing at the hyperbolic point $(q_2, p_2)$ is the separatrix of the big asymmetric double well. The red points indicate a chaotic sea that appears in the vicinity of the separatrix.
Around the minimum at $(0,0)$, the orbits are periodic and have frequencies $\Tilde{\omega}_0=\omega_0$.

\subsection{Double well at the phase space center}
Close to the stationary point $(q_0,p_0)=(0,0)$ at the center of the phase space, there is a bifurcation caused by the chosen driving frequency, $\omega_d \simeq 2 \omega_0$,  that gives rise to another double-well structure. This is better seen in Fig.~S~\ref{fig:SM02}(b), where we enlarge the area around $(q_0,p_0)$. The entire analysis developed in the main text concerns this region of the phase space.

The double-well structure in Fig.~S~\ref{fig:SM02}(b) also exhibits three critical points: two minima and a hyperbolic point. Notice that the hyperbolic point of this double well is very close to phase space center $(0,0)$. The separatrix is indicated with the red line, which corresponds to the Bernoulli lemniscate given by
\[  (q^2+p^2)^2=4 \Gamma(q^2-p^2),   
\]
and in polar coordinates 
\[
r^2(\theta)=4 \Gamma\cos(2\theta),
\]
 where the focal distance is $\sqrt{2 \Gamma}$. The surface area corresponds to 
\begin{equation}
    4\int^{\pi/4}_0  \int^{r(\theta)}_0 \!\! r\,dr\,d \theta = 4 \Gamma,
\end{equation}
which is the result used to obtain equation~(7) in the main text.

\begin{table}[h]
\begin{tabular*}{\textwidth}{@{\extracolsep\fill}cccccc}
\toprule%
Point & $10^{4}K/\omega_0$ & $\Gamma$ & $n_{\text{min}}$ & $\sqrt{2\Gamma}/|d_+|$ & $\sqrt{2}\Pi/|d_+|$  \\
\hline
A & 0.53 & 8.5 & 8.079 & 0.04122 & 0.00148492\\
B &  5.02 & 8.5 & 7.249 & 0.141397 & 0.0157244 \\
C & 0.53 & 80 & 77.007 & 0.12647 & 0.0140573\\
D & 2.91 & 80 & 66.134 & 0.29577 & 0.0769191\\
E & 8.33 & 80 & 197.924 & 0.49995 & 0.219769\\
F & 25 & 80 & 336.598 & 0.86594 & 0.659321\\  
\botrule
\end{tabular*}
\caption{This table is the same as Table 1 of the main text. It gives the parameters for the points A-F  marked in Fig.~1(g) of the main text, but now the values of $\sqrt{2 \Gamma}/|d_{+}|$ and $\sqrt{2} \Pi/|d_{+}|$ are also given. 
}
\label{tab:apppoints6}
\end{table}

In Fig.~S~\ref{fig:SM02}(c), $\sqrt{2}\Pi$ is the distance between the phase space center $(0,0)$  and the center (hyperbolic point) of the Bernoulli lemniscate. The separation between the two points can be understood as follows. The dynamics around the critical point $(0,0)$ is given by
\[
    q(t) = q_{0}(t) + q_r(t),
\]
where $q_{0}(t)$ is the homogeneous solution obtained with the undriven classical Hamiltonian $h_0$ and $q_r(t)$ is obtained from the linear terms of the Hamilton equations for the driven case, so that
\[
\ddot q_{r} + \omega^2_0 q_{r} = -\sqrt{2}\omega_d \Omega_d \sin \omega_d t,
\]
and 
\[
q_{r}(t)=\sqrt{2}\Pi \sin(\omega_d t), 
\]
where 
\[
\Pi=\Omega_d \omega_d/\left( \omega^2_d -\omega^2_0 \right).
\]
The linear response associated with $q_{r}(t)$ causes a translation of the center of the lemniscate by the amplitude $\sqrt{2}\Pi$.
Therefore, as one can see from Figs.~\ref{fig:SM02}(a)-(c) in the main text, the condition for the existence of a well-defined  inner double-well structure centered close to $(0,0)$ is  
\begin{equation}
|\sqrt{2}\Pi|+|\sqrt{2\Gamma}| < |d_+|.   
\label{eq:ineq}
\end{equation}

In Table~S~\ref{tab:apppoints6}, we complement Table 1 of the main text by providing  the values of $\sqrt{2 \Gamma}/|d_{+}|$ and $\sqrt{2} \Pi/|d_{+}|$.  All points, except for point F, satisfy the inequality in equation~(\ref{eq:ineq}). For point F, the lemniscate is already destroyed by chaos.

\section{Control parameters}
\label{app:controlParam}

\begin{figure*}[b]
    \bf{\centering
\includegraphics{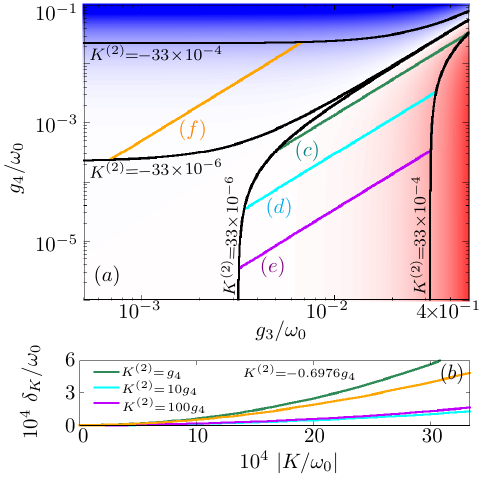}}
    \includegraphics[height=.5\textwidth,]{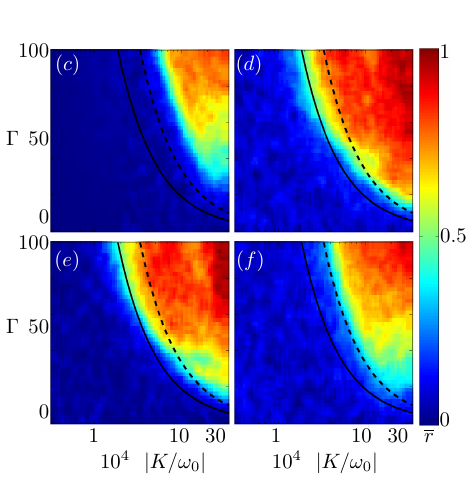}
    \caption{{\bf Parameter selection and quantum chaos maps.} {\bf a}{\normalfont, Kerr amplitude $K^{(2)}$ of the second-order effective Hamiltonian (in color) as a function of $g_3/\omega_0$ and $g_4/\omega_0$. Red is used for $K^{(2)}>0$ and blue for $K^{(2)}<0$; the solid black lines mark constant values of $K^{(2)}$; the green line marked as (c) is for $K^{(2)}=g_4$, the cyan line (d) is for $K^{(2)}=10g_4$, the purple line (e) is for $K^{(2)}=100g_4$, and the orange line (f) is for $K^{(2)}=-0.6976g_4$. {\bf b}, Absolute difference between $K^{(2)}$ and $K$ as a function of $|K/\omega_0|$ for different choices of $K^{(2)}=\mathcal{C}g_4$, as indicated. {\bf c-f}, Measure of quantum chaos given by the average ratio $\overline{r}$ of consecutive quasienergies spacings as a function of $K/\omega_0$ and $\Gamma$, for $\omega_d/\omega_0=1.999866$ and ({\bf c}) $K^{(2)}/g_4=1$, ({\bf d}) $K^{(2)}/g_4=10$, ({\bf e}) $K^{(2)}/g_4=100$, and ({\bf f}) $K^{(2)}/g_4=-0.6976$.  The  black solid curve  indicatesthe parametric case, where the classical Lyapunov exponent becomes positive in the vicinity of the center of the lemniscate, and the black dashed curve indicates the parameters for which chaos sets in both inside and outside the original lemniscate.} 
    }
    \label{fig:SM03}
\end{figure*}

In the main text, the values of $K/\omega_0$ are varied parametrically  by varying $g_3/\omega_0$ and $g_4/\omega_0$  according to the equation 
\begin{equation}
g_4 = \frac{20 g_3^2}{69 \omega_0}.
\label{eq:g4g3}
\end{equation}
This choice is made to guarantee that we reproduce the scenario in~\cite{FrattiniPrep}, where the second-order effective Hamiltonian describes very well the experiment. The second-order effective Hamiltonian is given by~\cite{FrattiniPrep},
\begin{equation}
\label{eq:HKC} \frac{\hat{H}^{(2)}_{\mathrm{eff}}}{\hbar} = - K^{(2)} \hat a^{\dagger2} \hat a^2 +\epsilon_2^{(2)} (\hat a^{\dagger2} + \hat a^{2}) ,
\end{equation}
where 
\begin{equation}
K^{(2)}= -\frac{3g_4}{2} +  \frac{10g_3^2}{3\omega_0} ,
\label{eq:K2}
\end{equation}
and $\epsilon_2^{(2)} = 2 g_3 \Omega_d/(3\omega_0)$. 
Equation~(\ref{eq:g4g3}) is the same as equation~(\ref{eq:K2}) when $K^{(2)}= 10  g_4$. In this section, we show what happens to the analysis in Fig.~1(g) of the main text for other choices of ${\cal C}$ in $K^{(2)}= {\cal C} \, g_4$.

In Fig.~S~\ref{fig:SM03}(a), we show in color the values of $K^{(2)}$  as a function of $g_3/\omega_0$ and $g_4/\omega_0$. Blue gradient is used when $K^{(2)}<0$ and red gradient for $K^{(2)}>0$. The green, cyan, purple, and orange lines indicate the examples where ${\cal C} =\{1,10,100,-0.6976\}$, respectively. In Fig.~S~\ref{fig:SM03}(b), we use the difference $\delta_K=|K-K^{(2)}|$ to compare $K^{(2)}$ and $K$. The behavior of $\delta_K$ with ${\cal C}$ is non-monotonic. The best match between $K$ and $K^{(2)}$ happens for  $K^{(2)}=10g_4$ (cyan line), which justifies the use of this choice for the analysis in the main text. 

We notice that for the experimental parameter $K/\omega_0=0.32/6000$ used in \cite{FrattiniPrep}, our choice of $K^{(2)}=10g_4$ implies that $g_3/\omega_0=25.7371/6000$, which is very close to the experimental value $g_3/\omega_0 = 30/6000$  used in that same work. The example  ${\cal C} = -0.6976$ is selected  also using the parameters $g_3/\omega_0=25.7371/6000$ and $K^{(2)}/\omega_0=-0.32/6000$, with the difference that $K^{(2)}$ is now negative. We investigate ${\cal C} = -0.6976$, because negative Kerr amplitudes are also experimentally available. 

Figure~S~\ref{fig:SM03}(d) is exactly the same as Fig.~1(g) of the main text. It shows the average value of the quantum chaos indicator $\bar{r}$ as a function of $\Gamma$ and $K/\omega_0$.
To complement the analysis of the regular to chaos transition performed in the main text,  we show in Fig.~S~\ref{fig:SM03}(c), Fig.~S~\ref{fig:SM03}(e), and Fig.~S~\ref{fig:SM03}(f) the results for $\bar{r}$ as a function of $\Gamma$ and $K/\omega_0$ for $K^{(2)}=g_4$, $K^{(2)}=100g_4$, and $K^{(2)}=-0.6976g_4$, respectively. The results are comparable,  although for  $K^{(2)}=g_4$ in Fig.~S~\ref{fig:SM03}(c), we see that the transition to chaos gets shifted to larger values of $\Gamma$ and $K/\omega_0$.

There are numerous ways in which the parameters of the Hamiltonian may be varied. There are various paths that can be taken to change $g_3$ and $g_4$ that are not necessarily linear, as those in Fig.~S~\ref{fig:SM03}, but the relationship in the equation~(12) of the main text is general. 


\end{document}